\documentclass[11pt, a4paper]{article}
\usepackage{jheparxiv}
\usepackage[latin1]{inputenc}
\usepackage{amsmath}
\usepackage{amsfonts}
\usepackage{amssymb}
\usepackage{latexsym}
\usepackage{mathrsfs}
\usepackage{graphicx}
\usepackage{color}
\usepackage{slashed}
\usepackage{twistor}

\subheader{\hfill \texttt{DAMTP-2015-14}}

\title{Scattering equations, supergravity integrands, and pure spinors}

\author{Tim Adamo and Eduardo Casali}

\affiliation{Department of Applied Mathematics \& Theoretical Physics \\
        University of Cambridge \\
        Wilberforce Road \\
        Cambridge CB3 0WA, United Kingdom}

\emailAdd{[t.adamo, e.casali]@damtp.cam.ac.uk}

\abstract{The tree-level S-matrix of type II supergravity can be computed in scattering equation form by correlators in a worldsheet theory analogous to a chiral, infinite tension limit of the pure spinor formalism. By defining a non-minimal version of this theory, we give a prescription for computing correlators on higher genus worldsheets which manifest space-time supersymmetry. These correlators are conjectured to provide the loop integrands of supergravity scattering amplitudes, supported on the scattering equations. We give non-trivial evidence in support of this conjecture at genus one and two with four external states. Throughout, we find a close correspondence with the pure spinor formalism of superstring theory, particularly regarding regulators and zero-mode counting.}

\begin{document}

\maketitle

\section{Introduction}

Recently, new expressions for the tree-level S-matrix of (the NS-NS sector of) supergravity have been found which are remarkably compact~\cite{Cachazo:2013hca}.\footnote{Similar representations have also been found for the tree-level S-matrices of a variety of other theories, including Yang-Mills, scalar, Dirac-Born-Infeld, and the non-linear sigma model~\cite{Cachazo:2013iea,Cachazo:2014xea}.} These formulae give the amplitudes in terms of an integral over the moduli space of a punctured sphere, with the moduli integrals localized on the support of the \emph{scattering equations}, which fix the punctures on the sphere in terms of the kinematic data. As their structure suggests, these new formulae arise from the genus zero correlation functions of vertex operators in a chiral, first order worldsheet theory, which resembles a chiral, infinite tension limit of the RNS formalism for type II superstring theory~\cite{Mason:2013sva}. The spectrum of this model contains only the massless states of type II supergravity, and it produces the tree-level S-matrix of the field theory \emph{exactly}, with no $\alpha^\prime$ or higher-derivative corrections. Indeed, formulating this RNS-like model on a curved background demonstrates that this worldsheet theory describes the NS-NS sector of supergravity at the non-linear level, with no higher-derivative corrections and the field equations emerging as quantum corrections to the scattering equations~\cite{Adamo:2014wea}.

Furthermore, on higher genus worldsheets, there is strong evidence that the correlation functions in the model correspond to the loop integrands of supergravity scattering amplitudes: they are modular invariant and factorize onto rational functions of kinematic data~\cite{Adamo:2013tsa}. At genus one and four points, they have even been shown to reproduce the correct kinematic prefactor and sum over scalar box integrals in the IR~\cite{Casali:2014hfa}. Much like the genus zero expressions, these more general amplitudes are localized on the support of the scattering equations at higher genus~\cite{Adamo:2013tsa}. Generally, these scattering equations are the minimal set of conditions needed to set a meromorphic quadratic differential equal to zero globally on the worldsheet. While type II supergravity is UV divergent in ten-dimensions, these divergences emerge only after integrating over a set of zero-modes in the worldsheet theory which correspond to loop integrals in field theory; the remaining integrand is modular invariant and has the expected factorization properties. However, space-time supersymmetry is manifested only after summing over worldsheet spin structures, just as in the RNS formalism for superstrings.

\medskip

In the context of superstring theory, the pure spinor formalism gives a manifestly super-Poincar\'e-invariant quantization which avoids the difficulties of dealing with space-time supersymmetry or light-cone gauge in the RNS and Green-Schwarz formalisms, respectively~\cite{Berkovits:2000fe,Berkovits:2002zk,Bedoya:2009np}. By now, this pure spinor formalism has been used extensively in the study of perturbative scattering amplitudes, enabling explicit calculations at higher-genus which have so far been beyond the reach of other methods (\textit{e.g.}, \cite{Gomez:2013sla}).

A pure spinor version of the chiral, `infinite tension' worldsheet model has also been proposed~\cite{Berkovits:2013xba}, and shown to give the correct tree-level S-matrix of fully supersymmetric type II supergravity~\cite{Gomez:2013wza}. Given the efficacy of the pure spinor approach to superstring amplitudes at higher genus, it seems natural to ask if there is a prescription for the calculation of loop integrands in supergravity using this formalism.

In this paper, we provide a potential answer to this question by developing the pure spinor worldsheet theory for supergravity at higher genus. In the context of the pure spinor superstring, the higher-genus amplitude prescription is best realized by a \emph{non-minimal} extension of the worldsheet model (\textit{c.f.}, \cite{Berkovits:2005bt,Berkovits:2006vi}). Consequently, we define a non-minimal version of the supergravity worldsheet model in section \ref{sec:nonmin}, giving the worldsheet action, BRST charge, effective $b$-ghost, and regulator prescriptions. In many aspects, these objects closely resemble (or are even identical to) their string theoretic counterparts, while also inheriting much of the structure of pure spinor world\emph{line} formalisms for supergravity~\cite{Berkovits:2001rb,Bjornsson:2010wu}.

Section \ref{sec:loop} explores the resulting amplitude prescription at arbitrary genus; we conjecture that the genus $g$ worldsheet correlator gives the integrand of the $g$-loop supergravity amplitude; UV divergences arise from integrating over non-compact zero-modes in the model. The general structure of these correlators, and the associated scattering equations, is explored. At tree-level, the amplitudes reproduce the `minimal' prescription of~\cite{Berkovits:2013xba} which is known to give the tree-level S-matrix of type II supergravity~\cite{Gomez:2013wza}. At genus one and two, the four-point function passes several checks in favor of our conjecture, including producing the correct supersymmetric prefactors and factorizing like a field theory amplitude. Section \ref{sec:concl} concludes.

Appendix \ref{app:currents} catalogs the various worldsheet currents that appear through the paper and their OPEs. In appendix \ref{app:gf}, we note that the worldsheet model studied in this paper can be obtained from a gauge-fixing procedure, in direct analogy with a recent proposal for the superstring~\cite{Berkovits:2014aia}.


\section{Non-minimal formalism}
\label{sec:nonmin}

In~\cite{Berkovits:2013xba}, a worldsheet theory akin to a chiral, infinite tension limit of the pure spinor superstring was proposed. The vertex operators of this model encode the full type II supergravity multiplet, and the genus zero correlation functions give the tree-level S-matrix of supergravity~\cite{Gomez:2013wza}. To explore the meaning of correlation functions on higher genus worldsheets in this model, we must first provide a sensible amplitude prescription at generic genus.

In the superstring context, a tentative higher genus prescription can be made using the `minimal' worldsheet variables; unfortunately, it entails the use of complicated picture changing operators to define the functional integrals in play~\cite{Berkovits:2004px}. Furthermore, the prescription for integrating over the worldsheet modular parameters requires an effective $b$-antighost which is not manifestly covariant (\textit{i.e.}, the definition depends on the choice of a patch of pure spinor space)~\cite{Berkovits:2001us}. While explicit calculations at genus one~\cite{Berkovits:2004px,Mafra:2005jh} and two~\cite{Berkovits:2005df,Berkovits:2005ng} can be made with this formalism, the picture changing operators complicate the functional integration and break manifest Lorentz covariance at intermediate stages, although the final amplitudes are covariant (\textit{c.f.}, \cite{Mafra:2009wq}).

A more elegant prescription is provided by adding \emph{non-minimal} worldsheet variables to the model and modifying the BRST charge~\cite{Berkovits:2005bt,Berkovits:2006vi}. This eliminates the need for picture changing operators and allows one to define a covariant effective $b$-ghost to perform moduli integrals. Following this analogy with the superstring, we develop a non-minimal version of the pure spinor formalism for type II supergravity in order to enable higher genus calculations. In the next section, this non-minimal formalism will be used to provide an explicit amplitude prescription.


\subsection{Review of the minimal theory}

We begin with a brief review of the `minimal' pure spinor formalism proposed for supergravity in~\cite{Berkovits:2013xba}. Throughout, $m,n,\ldots =0,\ldots,9$ are ten-dimensional space-time indices, while $\alpha,\beta,\ldots =1,\ldots,16$ are spinor indices. The model has a chiral, first order action given by a holomorphic complexification of the pure spinor superparticle action~\cite{Berkovits:2001rb}:
\be\label{min1}
S=\frac{1}{2\pi}\int_{\Sigma} P_{m}\,\dbar X^{m}+p_{\alpha}\,\dbar\theta^{\alpha}+\tilde{p}_{\til\alpha}\,\dbar\til\theta^{\til\alpha} +w_{\alpha}\,\dbar\lambda^{\alpha}+\tilde{w}_{\til\alpha}\,\dbar\til{\lambda}^{\til\alpha}\,
\ee
where $X^{m}$ and $\theta^{\alpha}, \tilde{\theta}^{\til\alpha}$ are the bosonic and fermionic matter variables, while $P_{m}$ and $p_{\alpha}, \tilde{p}_{\til\alpha}$ are their conjugate momenta, which have holomorphic worldsheet conformal weight $(1,0)$. 

The variables $\lambda^{\alpha},\tilde{\lambda}^{\til\alpha}$ are bosonic spinors which satisfy an algebraic purity condition:
\begin{equation}\label{pscond}
 \lambda^{\alpha}\gamma^{m}_{\alpha\beta}\lambda^{\beta}=0=\tilde{\lambda}^{\til\alpha}\gamma^{m}_{\til\alpha\til\beta}\tilde{\lambda}^{\til\beta}\,,
\end{equation}
while $w_{\alpha},\tilde{w}_{\til\alpha}$ are their conformal weight $(1,0)$ conjugate momenta. The pure spinor condition reduces the number of independent components of $\lambda, \tilde{\lambda}$ from sixteen to eleven, and ensures that their conjugates can only ever appear in currents which are invariant under the induced gauge symmetry:
\begin{equation*}
 N^{nm}=\frac{1}{2}(w\gamma^{nm}\lambda)\,, \qquad J=\lambda\cdot w\,, \qquad T_{\lambda}=-w_{\alpha}\,\partial \lambda^{\alpha}
\end{equation*}
The pure spinor constraint also ensures that the theory has vanishing conformal anomaly, confirming the choice of critical dimension $d=10$.

Since the action \eqref{min1} is free and first-order, the OPEs between the matter variables are simply
\be\label{matOPE}
X^{m}(z)\,P_{n}(w)\sim \frac{\delta^{m}_{n}}{z-w}\,, \qquad \theta^{\alpha}(z)\,p_{\beta}(w)\sim\frac{\delta^{\alpha}_{\beta}}{z-w}\,,
\ee
and likewise for the tilded variables. The OPEs between the various currents of pure spinor variables can be computed in the same fashion as the superstring (\textit{e.g.}, by working with the $\U(5)$-covariant parametrization of the space of pure spinors)~\cite{Berkovits:2000fe}, and are collected in appendix \ref{app:currents} for reference.

The BRST operator for this model corresponds to a holomorphic generalization of the same operator in the type II worldline formalism. In particular, defining the Green-Schwarz constraint
\be\label{GScon}
d_{\alpha}=p_{\alpha}-\frac{1}{2}P_{m}\gamma^{m}_{\alpha\beta}\theta^{\beta}\,,
\ee
the charge
\be\label{minBRST}
Q=\oint \lambda^{\alpha}\,d_{\alpha}+\til{\lambda}^{\til\alpha}\,\til{d}_{\til\alpha}\,,
\ee
is easily seen to be nilpotent, using the OPEs \eqref{matOPE} and the pure spinor conditions \eqref{pscond} on $\lambda,\til\lambda$. Vertex operators are given by non-trivial cohomology classes with respect to this BRST operator. The fixed vertex operator is equal to that of the superparticle; in a momentum eigenstate representation:
\be\label{fVO}
V=\lambda^{\alpha}\,\til{\lambda}^{\til\alpha}A_{\alpha}(\theta)\,\tilde{A}_{\til\alpha}(\til{\theta})\,\e^{i k\cdot X}\,,
\ee
where $A_{\alpha},\tilde{A}_{\til\alpha}$ are the standard $\cN=1$ superfields, which can be expanded in terms of vector and spinor polarizations. The condition $\{Q,V\}=0$ enforces the linearized equations of motion
\begin{equation*}
 k^2=0\,, \qquad (\gamma_{mnpqr})^{\alpha\beta}D_{\alpha}A_{\beta}=0=(\gamma_{mnpqr})^{\til\alpha \til\beta}\tilde{D}_{\til\alpha}\tilde{A}_{\til\beta}\,,
\end{equation*}
where the supersymmetric derivative is the usual
\begin{equation*}
 D_{\alpha}=\frac{\partial}{\partial\theta^\alpha}+\frac{1}{2}k^{m} (\gamma_{m}\theta)_{\alpha}\,.
\end{equation*}

The integrated vertex operators resemble those of the type II pure spinor superstring, but the chirality of the model leads to the presence of some holomorphic delta functions to balance the conformal weight\footnote{Notice also the absence of terms proportional to $\partial\theta$ and $\partial\tilde\theta$.}:
\begin{multline}\label{iVO}
\int_{\Sigma}\bar{\delta}(k\cdot P)\,U \\
=\int_{\Sigma}\bar{\delta}(k\cdot P)\,\left(A\cdot P+ d_{\alpha}W^{\alpha}+\frac{1}{2}N_{mn}\mathcal{F}^{mn}\right) \left(\tilde{A}\cdot P+\tilde{d}_{\til\alpha}\tilde{W}^{\til\alpha}+\frac{1}{2}\tilde{N}_{mn}\tilde{\mathcal{F}}^{mn}\right)\,\e^{ik\cdot X}\,,
\end{multline}
where $\{A^{m},W^{\alpha},\mathcal{F}^{mn},\ldots\}$ are the standard superfields of $\cN=1$ super-Yang-Mills in ten dimensions (\textit{c.f.}, \cite{Harnad:1985bc,Witten:1985nt}).  This operator also obeys $[Q,U]=0$ on the support of the delta function.\footnote{While the delta function is included `by hand' in this formalism to give appropriate conformal weights and BRST closure of the vertex operator, its presence can be understood via worldsheet gauge-fixing in the RNS-like version of this model~\cite{Adamo:2013tsa}.} 

The vertex operators \eqref{fVO}, \eqref{iVO} give the full spectrum of type II supergravity in ten dimensions. Individual fields can be picked out by expanding the various superfields in powers of $\theta$ (or $\til\theta$), and isolating those components proportional to the desired polarizations.

The genus zero worldsheet correlation function prescription given in~\cite{Berkovits:2013xba} mimics the prescription for the superstring:
\be\label{minamp1}
\cM_{n}^{(0)}=\left\la \prod_{i=1}^{3}V(z_i)\,\prod_{j=4}^{n}\int_{\Sigma}\bar{\delta}(k_i\cdot P(z_i))\,U(z_i)\right\ra\,,
\ee
with the usual zero-mode normalization for $\theta,\tilde{\theta},\lambda,\tilde{\lambda}$ inherited from the superstring (\textit{i.e.}, $\la\lambda^3 \theta^5\ra=1$)~\cite{Berkovits:2000fe}.  By restricting the vertex operators to the NS-NS sector, it is straightforward to see that this prescription reproduces the worldsheet correlators of the RNS-like model in~\cite{Mason:2013sva}. These in turn are equal to the scattering equation representations for the tree-level S-matrix of gravitons, $B$-fields, and dilatons given by Cachazo, He, and Yuan~\cite{Cachazo:2013hca}.

In the case of general supersymmetric external states, performing explicit amplitude calculations for an arbitrary number of external particles is difficult. However, by utilizing genus zero results from the pure spinor superstring~\cite{Mafra:2010jq,Mafra:2011nv,Broedel:2013tta} and KLT orthogonality, it can nevertheless be shown that the prescription \eqref{minamp1} \emph{does} reproduce the full tree-level S-matrix of type II supergravity, in a representation that is supported on the scattering equations~\cite{Gomez:2013wza}. Note that the distinction between type IIA and IIB supergravity is built into the identification of the tilded spinor indices: for IIA tilded indices denote spinors of the \emph{opposite} chirality as un-tilded indices, while for IIB they denote spinors of the \emph{same} chirality.


\subsection{Non-minimal worldsheet action and BRST charge}

Following the analogy with superstring theory, we define the non-minimal version of the model \eqref{min1} by the inclusion of two sets of new variables on the worldsheet: bosonic spinors $\bar{\lambda}_{\alpha}, \til{\bar{\lambda}}_{\til\alpha}$ and fermionic spinors $r_{\alpha}, \tilde{r}_{\til\alpha}$, along with their respective conjugate fields $\bar{w}^{\alpha}, \til{\bar{w}}_{\til\alpha}$ and $s^{\alpha}, \tilde{s}^{\til\alpha}$.  These variables obey the constraints
\begin{equation}
 \bar{\lambda}_{\alpha}\gamma_{m}^{\alpha\beta}\bar{\lambda}_{\beta}=0=\til{\bar{\lambda}}_{\til\alpha}\gamma_{m}^{\til\alpha\til\beta}\til{\bar{\lambda}}_{\til\beta}\,, \qquad \bar{\lambda}_{\alpha}\gamma_{m}^{\alpha\beta}r_{\beta}=0=\til{\bar{\lambda}}_{\til\alpha}\gamma_{m}^{\til\alpha\til\beta}\tilde{r}_{\til\beta}\,.
\end{equation}
This means that $\bar{\lambda},\til{\bar\lambda}$ are pure spinors of opposite chirality to $\lambda, \til\lambda$; if the space-time signature is taken to be Euclidean, then they can be interpreted as complex conjugates of the original variables. The constraints also restrict the fermions $r,\tilde{r}$ to having eleven independent components.

The modified action is now
\be\label{nmact}
S=\frac{1}{2\pi}\int_{\Sigma} P_{m}\,\dbar X^{m}+p_{\alpha}\,\dbar\theta^{\alpha}+w_{\alpha}\,\dbar\lambda^{\alpha}+\bar{w}^{\alpha}\,\dbar\bar{\lambda}_{\alpha}+s^{\alpha}\,\dbar r_{\alpha} \: +\,\mbox{tilded}\,.
\ee
The constraints ensure that the $\bar{w}\bar{\lambda}$-system contributes $+22$ units of central charge, which is balanced by the $-22$ contributions from the $rs$-system. Hence, this non-minimal worldsheet action has vanishing conformal anomaly for space-time dimension $d=10$, just like the minimal version. The action for the non-minimal fields is free, but the various constraints require a careful treatment of their OPEs. In particular, the variables of conformal weight $(1,0)$ can only appear in currents that are invariant under the gauge transformations induced by the pure spinor constraints. These are precisely the same as those used in the superstring~\cite{Berkovits:2005bt}:
\begin{equation*}
 \bar{N}_{mn}=\frac{1}{2}\left(\bar{w}\gamma_{mn}\bar{\lambda}+s\gamma_{mn}r\right)\,, \qquad \bar{J}=\bar{w}\cdot\bar{\lambda}+s\cdot r\,, \qquad T_{\bar{\lambda},r}=-\bar{w}^{\alpha}\partial\bar{\lambda}_{\alpha}-s^{\alpha}\partial r_{\alpha}\,,
\end{equation*}
\begin{equation*}
S_{mn}=\frac{1}{2}(s\gamma_{mn}\bar{\lambda})\,, \qquad S=s\cdot\bar{\lambda}\,.
\end{equation*}
The currents for the tilded variables are identical, and have the \emph{same} conformal weight as the un-tilded currents. The various OPEs between these currents are collected in appendix \ref{app:currents} for reference.

We define the non-minimal BRST operator to be
\be\label{nmBRST}
Q=\oint \lambda^{\alpha}d_{\alpha}+\tilde{\lambda}^{\til\alpha}\tilde{d}_{\til\alpha}+\bar{w}^{\alpha}r_{\alpha}+\tilde{\bar{w}}^{\til\alpha}\tilde{r}_{\til\alpha}\,,
\ee
which is nilpotent due to the pure spinor constraint. Since the `quartet' of non-minimal variables do not affect $Q^2=0$, standard arguments~\cite{Kugo:1979gm,Rybkin:1989ms} ensure that they have no impact on the BRST cohomology. In particular, external supergravity states can be represented in the non-minimal worldsheet model by the same fixed \eqref{fVO} and integrated \eqref{iVO} vertex operators used in the minimal model.

Note that just as the minimal model and BRST charge \eqref{min1}, \eqref{minBRST} resemble a holomorphic complexification of the pure spinor superparticle, the non-minimal action and BRST charge \eqref{nmact}, \eqref{nmBRST} are a holomorphic complexification of the non-minimal superparticle developed in~\cite{Bjornsson:2010wu}. This worldline formalism has been used to check the UV divergence structure of maximally supersymmetric supergravity loop amplitudes~\cite{Bjornsson:2010wm}, suggesting that the worldsheet model should be related to field theory beyond tree-level.


\subsection{Effective $b$-ghost}

In the RNS formalism for superstring theory, the prescription for integrating over worldsheet moduli at arbitrary genus is provided by the functional integral over the conformal $bc$-ghost system. In the RNS-like worldsheet formulation of supergravity, there are \emph{two} conformal ghost systems: one corresponds to gauging the worldsheet stress tensor as in string theory, while the other corresponds to gauging the Hamiltonian constraint $P^2=0$~\cite{Mason:2013sva}. This latter constraint ensures that the resulting worldsheet correlation functions are supported on the scattering equations -- indeed, in the presence of vertex operator insertions, $P^2=0$ is \emph{equivalent} to the scattering equations at any genus~\cite{Adamo:2013tsa}.

Of course, there is no $bc$-ghost system in either the pure spinor superstring or the worldsheet model discussed here. In the superstring, a prescription for integrating over moduli is nonetheless available by defining a composite operator $b\in\Pi\Omega^{0}(\Sigma, K^2_{\Sigma})$, called an effective $b$-ghost, which obeys $\{Q,b\}=T$. In our model, it is also possible to construct an effective $b$-ghost, but instead of being related to the stress tensor, this composite operator obeys $\{Q,b\}=P^2$. The effective $b$-ghost of the pure spinor superparticle is also related to the Hamiltonian constraint (albeit a real function on the worldline rather than a quadratic differential on the worldsheet), and ensures the gauge invariance of the propagator~\cite{Bjornsson:2010wu}. Viewing our formalism as a complexification of the worldline theory, this choice of ghost will likewise ensure gauge invariance, as well as modular invariance and the appropriate scattering equations at arbitrary genus.   

While the lack of an explicit relationship with the stress tensor is slightly mysterious, it seems to be related to the fact that both the Virasoro and Hamiltonian constraints are implied by a single twistor-like constraint in conjunction with a $\lambda^{\alpha}$ constraint (see appendix \ref{app:gf}). In the superstring, the twistor-like constraint implies the Virasoro constraint only~\cite{Berkovits:2014aia}. Of course, the ultimate test of our choice will be the resulting amplitude prescription. 

\medskip

The construction of the effective $b$-ghost proceeds in direct analogy to the superstring calculation (\textit{c.f.}, \cite{Berkovits:2005bt,Oda:2007ak}). We begin by looking for an operator $G^{\alpha}\in\Pi\Omega^{0}(\Sigma, K^{2}_{\Sigma})$ which obeys $\{Q,G^{\alpha}\}=\lambda^{\alpha}P^2$. Using the various OPEs between currents and fields in our worldsheet model, it is easy to see that
\begin{equation*}
 G^{\alpha}=-P_{m}\,(\gamma_{m}d)^{\alpha}\,,
\end{equation*}
has the desired property. Now, since 
\begin{equation*}
 \{Q, \lambda^{\alpha} G^{\beta}\}=\lambda^{\alpha}\lambda^{\beta}\,P^2\,,
\end{equation*}
the operator $(\lambda^{\alpha} G^{\beta}-\lambda^{((\alpha} G^{\beta))})$ is BRST-closed, where $((\cdots))$ denotes the symmetric, gamma-matrix-traceless part. As the $Q$-cohomology at ghost number one with non-zero conformal weight is trivial, there must exist some $H^{\alpha\beta}$ of conformal weight $(2,0)$ such that 
\begin{equation*}
 \left[Q, H^{\alpha\beta}-H^{((\alpha\beta))}\right]=\lambda^{\alpha} G^{\beta}-\lambda^{((\alpha} G^{\beta))}\,.
\end{equation*}
A calculation identical to the analogous step in the superstring reveals that
\begin{equation*}
 H^{\alpha\beta}=\frac{(\gamma^{mnp})^{\alpha\beta}}{96}\left[(d\gamma_{mnp}d)+24 N_{mn}P_{p}\right]\,.
\end{equation*}

Cohomological arguments allow for the continued construction of a chain of operators, each related to the previous operator in the chain by the action of $Q$, until the chain terminates by virtue of the pure spinor constraint. These operators can then be arranged into a single composite operator by making use of the non-minimal pure spinor variables:
\begin{multline}\label{effbg}
 b=-\frac{(\bar{\lambda}\gamma^{m}d)P_{m}}{\bar{\lambda}\cdot\lambda}-\frac{(\bar{\lambda}\gamma^{mnp}r)}{96 (\bar{\lambda}\cdot\lambda)^2}\left[(d\gamma_{mnp}d)+24 N_{mn}P_{p}\right] \\ + \frac{(r\gamma_{mnp}r)(\bar{\lambda}\gamma^{m}d)}{8 (\bar{\lambda}\cdot\lambda)^3} N^{np} -\frac{(r\gamma_{mnp}r)(\bar{\lambda}\gamma^{pqr}r)}{64 (\bar{\lambda}\cdot\lambda)^4}N^{mn} N_{qr}\,,
\end{multline}
which obeys $\{Q,b\}=P^2$. The effective $b$-ghost for the tilded worldsheet fields takes an identical form. This composite operator is identical to the holomorphic complexification of the $b$-ghost appearing in the non-minimal pure spinor superparticle~\cite{Bjornsson:2010wu,Bjornsson:2010wm}, up to an overall constant factor.


\subsection{Zero modes, functional integrals, and regulators}

In any path integral calculation, regardless of the details of the amplitude prescription, zero modes of the various worldsheet fields must be integrated over. Remarkably, the only variable in the model \eqref{nmact} which does not appear in the pure spinor superstring is $P_{m}\in\Omega^{0}(\Sigma,K_{\Sigma})$; all other worldsheet fields appear as left-movers in the superstring. Hence, the subtleties associated with their functional integrations can be dealt with in \emph{exactly} the same manner as they are handled in the superstring context. Crucially, the tilded sector of the worldsheet model is just a second (left-moving) copy of the un-tilded sector.

The conformal weight zero matter fields $\theta^{\alpha}, \tilde{\theta}^{\til\alpha}$ have the usual zero mode integration measures, which will be denoted by $\d^{16}\theta$, $\d^{16}\til\theta$ at arbitrary worldsheet genus. Likewise, at any genus the bosonic and fermionic pure spinor variables $\lambda^{\alpha}$, $\bar{\lambda}_{\alpha}$, $r_{\alpha}$ and their tilded counterparts have eleven zero modes. Since these are identical to the pure spinor variables of the superstring, we can use the same integration measures that were developed in that context for both the tilded and un-tilded variables. The precise definition of the zero mode measures can be found in~\cite{Berkovits:2004px,Berkovits:2004bw,Berkovits:2005bt}; we will simply denote them as $[\d\lambda]$, $[\d\bar{\lambda}]$, $[\d r]$, etc.

All of the conjugate fields in this model are left-moving, with conformal weight $(1,0)$. So on a genus $g$ worldsheet, they acquire $g$ zero modes which must be integrated over. Let $f$ be any such worldsheet field; at genus $g$ we expand it as
\begin{equation}
 f\rightarrow \widehat{f}+\sum_{I=1}^{g}f^{I}_{\mathrm{z.m.}}\,\omega_{I}\,,
\end{equation}
where $\widehat{f}$ is the quantum (non-zero mode) field, $\{\omega_{I}\}$ form a basis of $H^{0}(\Sigma,K_{\Sigma})$, and $f^{I}_{\mathrm{z.m.}}$ are the functions (bosonic or fermionic) which parametrize the zero modes. Choosing a canonical basis $\{A_{1},\ldots,A_{g},B_{1}\ldots,B_{g}\}$ for $H_{1}(\Sigma, \Z)\cong \Z^{2g}$ and the $\{\omega_{I}\}$ such that
\begin{equation*}
 \int_{A_I}\omega_{J}=\delta_{IJ}\,, \qquad \int_{B_{I}}\omega_{J}=\Omega_{IJ}\,,
\end{equation*}
where $\Omega_{IJ}$ is the period matrix of $\Sigma$, the zero mode of a field can be extracted unambiguously as
\begin{equation*}
 f^{I}_{\mathrm{z.m.}}=\int_{A_I} f\,.
\end{equation*}
The various conformal weight one fields which have zero mode structure of this form are $p_{\alpha}, w_{\alpha}, \bar{w}^{\alpha}, s^{\alpha}$ and their tilded counterparts (we leave the field $P_{m}$ to be treated later). Once again, their zero mode integrals can be performed in an identical manner to the left-movers of the superstring. Hence, we trade the integral over $p^{I}_{\mathrm{z.m.}\,\alpha}$ for an integral over $d_{\mathrm{z.m.}\,\alpha}^{I}$, and denote the various integral measures by
\begin{equation}
 [\d d]=\prod_{I=1}^{g}[\d^{16}d^{I}_{\mathrm{z.m.}}]\,, \qquad [\d w]=\prod_{I=1}^{g}[\d^{11}w^{I}_{\mathrm{z.m.}}]\,, \; \ldots
\end{equation}
The technical definitions of these measures can be found throughout the literature on the pure spinor formalism~\cite{Berkovits:2004px,Berkovits:2004bw,Berkovits:2005bt}.

Just as in the pure spinor superstring, there are two important subtleties associated with these zero mode integrations. Firstly, there are non-compact integrals which can introduce potential divergences. If the non-minimal formalism is to be equivalent to the minimal prescription at genus zero, it cannot have new divergences, so these integrals require regularization. But since the pure spinor variables of our worldsheet model are identical to the left-moving pure spinor variables of the superstring, we can use the \emph{same} regulator.  In particular, inserting $\cN=\exp(\{Q,\chi\})$ will not affect worldsheet correlation functions of BRST-closed vertex operators, so on a genus $g$ worldsheet we set~\cite{Berkovits:2005bt}
\be\label{reg1}
\chi=-\bar{\lambda}\cdot\theta -\sum_{I=1}^{g}\left(N^{I}_{\mathrm{z.m.}\,mn}\,S^{mn\,I}_{\mathrm{z.m.}}+J^{I}_{\mathrm{z.m.}}\,S^{I}_{\mathrm{z.m.}}\right)\,.
\ee
The exponential suppression then provides a regulator for the large $\lambda, \bar{\lambda}$ region.

The second subtlety arises from the zero mode integration near the tip of the pure spinor cone, where $\bar{\lambda}\cdot\lambda\rightarrow 0$. It can be shown that the zero mode measures are convergent in this region~\cite{Berkovits:2004bw,Berkovits:2005bt}:
\begin{equation*}
 [\d\lambda]\,[\d\bar{\lambda}]\,[\d r]\sim \lambda^{8} \bar{\lambda}^{11}\,.
\end{equation*}
However, the effective $b$-ghost \eqref{effbg} contains a term which diverges like $(\bar{\lambda}\cdot\lambda)^{-3}$ near the tip of the pure spinor cone. We expect to insert $3g-3$ such $b$-ghosts for any correlator on a genus $g\geq2$ worldsheet, so potential divergences can arise for $g>2$.  

Once more, we can look to the superstring context for a resolution for this problem. In that setting, a solution has been proposed in the form of a BRST-invariant regularization of the effective $b$-ghost. While the functional form of \eqref{effbg} differs slightly from the effective $b$-ghost of the superstring, its dependence on the pure spinor variables is the same, so we can adopt the pure spinor regularization for the $b$-ghost given by Berkovits and Nekrasov~\cite{Berkovits:2006vi}. The precise details of this regularization will not be needed for our considerations here, since we already know that it is designed to resolve the issue of divergences in the functional integration near the tip of the pure spinor cone.

The regularized $b$-ghost will be denoted by $b_{\epsilon}$; accounts of its use in several calculations can be found in~\cite{Berkovits:2006vi,Aisaka:2009yp,Grassi:2009fe}. We note that this prescription has yet to be used in a full, non-trivial superstring amplitude computation (the divergences do not arise for the four-point function until $g\geq4$ due to fermionic zero mode saturation). However, any potential issues which could arise from practical computations in the superstring will be identical in our worldsheet model.  


\section{Amplitudes at Higher Genus}
\label{sec:loop}

We are now in a position to define the amplitude prescription on a worldsheet of generic genus. We conjecture that the resulting amplitudes on a genus $g$ worldsheet contain the $g$-loop integrand of type II supergravity, with UV divergences emerging only after integrating over a non-compact space of zero modes corresponding to loop momenta. We can make many general observations about the structure of these amplitudes with an arbitrary number of external points, and provide concrete, non-trivial evidence in favor of the conjecture at four-points. In particular, for $g=1,2$ the four-point function has IR behavior consistent with supergravity and gives the correct supersymmetric prefactor. 


\subsection{Amplitude prescription}

Given the similarities between the model \eqref{nmact} and the superstring, we follow~\cite{Berkovits:2005bt} in giving the higher genus amplitude prescription.  In particular, on a genus $g\geq 2$ worldsheet we define the $n$-point amplitude by the worldsheet correlation function:
\be\label{ampg}
\cM_{n}^{(g)}= \lim_{\epsilon\rightarrow0}\int \prod_{a=1}^{3g-3}\d \tau_{a}\,\left\la \cN \widetilde{\cN} \prod_{j=1}^{3g-3}\bar{\delta}\left(P^2(z_j)\right)\,(b_{\epsilon}|\mu)_{j} (\tilde{b}_{\epsilon}|\tilde{\mu})_{j} \prod_{i=1}^{n}\int_{\Sigma}\bar{\delta}\left(k_{i}\cdot P(z_i)\right)\,U(z_i)\right\ra\,.
\ee
The complex parameters $\{\tau_{a}\}$ are the complex structure moduli of the genus $g$ Riemann surface $\Sigma$ integrated over the fundamental domain of the modular group\footnote{This is consistent with modular invariance. As usual for the pure spinor formalism, modular invariance is somewhat obscure at the level of the correlation function and only becomes manifest in the final amplitude. Our explicit four-point calculations confirm this.}; $\cN,\widetilde{\cN}$ are the regulators defined by \eqref{reg1}; $b,\tilde{b}$ are the effective $b$-ghosts of \eqref{effbg}; $\epsilon$ is the regulation parameter of~\cite{Berkovits:2006vi}; and $U(z)$ is the integrated vertex operator \eqref{iVO}. The Beltrami differentials $\mu_{j}$ form a basis of $H^{0,1}(\Sigma, T_{\Sigma})$, with
\begin{equation*}
 (b|\mu):=\,\frac{1}{2\pi} \int_{\Sigma}\mu\lrcorner b\,,
\end{equation*}
and likewise for the tilded variables. The brackets $\la\cdots\ra$ indicate the correlator in the worldsheet CFT; that is, integrating over zero modes and eliminating non-zero modes via worldsheet OPEs. Note that for $g=2$, the regulator $\epsilon$ can be dropped from this prescription.

As usual, the amplitude prescription for a genus one worldsheet should include a single fixed vertex operator in accordance with the constant translation symmetry on the torus or ghost number anomaly. Thus, the $g=1$ amplitudes are defined by
\be\label{amp1}
\cM_{n}^{(1)}=\int \d\tau\,\left\la \cN\widetilde{\cN}\,\bar{\delta}\left(P^2(z_1)\right) (b|\mu) (\tilde{b}|\tilde{\mu})\,V(z_1)\,\prod_{i=2}^{n}\int_{\Sigma}\bar{\delta}\left(k_{i}\cdot P(z_i)\right)\,U(z_i)\right\ra\,.
\ee
On the Riemann sphere, we have three fixed vertex operators in accordance with $\SL(2,\C)$ invariance, leading to
\be\label{amp0}
\cM_{n}^{(0)}=\left\la \cN\widetilde{\cN}\,\prod_{i=1}^{3}V(z_i)\,\prod_{j=4}^{n}\int_{\Sigma}\bar{\delta}(k_i\cdot P(z_i))\,U(z_i)\right\ra\,.
\ee

Despite the apparent complexity of the general amplitude prescription, there are nevertheless some important universal properties which can be easily observed. Note that with the momentum eigenstates of \eqref{fVO}, \eqref{iVO}, the worldsheet field $X^m$ enters the correlator only via the plane wave exponentials $\e^{i k\cdot X}$. Following the strategy adopted for the RNS-like model~\cite{Mason:2013sva}, the $X$ path integral can be performed explicitly, enforcing ten-dimensional momentum conservation and the equation of motion
\be\label{Peom}
\dbar\,P_{m}(z)=2\pi i\,\d z \wedge \d\bar{z}\,\sum_{i=1}^{n}k_{i\;m} \delta^{2}(z-z_i)\,.
\ee
This indicates that $P_m$ is a meromorphic differential on $\Sigma$, with singularities only at the vertex operator insertions $\{z_i\}\subset\Sigma$. 

On a genus $g$ Riemann surface, the kernel of $\dbar:\cO\rightarrow K_{\Sigma}$, denoted by $\tilde{S}_{g}(z,w|\Omega)$, serves as the propagator for the $PX$-system. This is a $(1,0)$-form with respect to $z$ and a scalar with respect to $w$, and can be defined as
\begin{equation*}
 \tilde{S}_{g}(z,w|\Omega)=\partial_{z}\,G_{g}(z,w|\Omega)\,,
\end{equation*}
\be\label{PXprop}
G_{g}(z,w|\Omega)=-\ln|E_{g}(z,w)|^2+2\pi \sum_{I,J=1}^{g}\left(\mathrm{Im}\,\Omega\right)^{-1}_{IJ}\left(\mathrm{Im}\int_{z}^{w}\omega_{I}\right)\left(\mathrm{Im}\int_{z}^{w}\omega_{J}\right)\,,
\ee
where $E_{g}(z,w)$ is the prime form (\textit{c.f.}, \cite{Fay:1973,D'Hoker:1988ta}). In the limit where $z\rightarrow w$, this propagator has the expected simple pole
\begin{equation*}
 \lim_{z\rightarrow w}\tilde{S}_{g}(z,w|\Omega)\sim \frac{\d z}{z-w}\,,
\end{equation*}
in appropriately chosen inhomogeneous coordinates on $\Sigma$.

Using \eqref{PXprop}, we can integrate \eqref{Peom} on the worldsheet, finding
\be\label{gP}
P_{m}(z)=\sum_{I=1}^{g}\ell^{I}_{m}\,\omega_{I}(z)+\sum_{i=1}^{n}k_{i\;m}\,\tilde{S}_{g}(z,z_i|\Omega)\,.
\ee
Combined with the on-shellness of the $\{k_i\}$, this indicates that $P^2$ is a meromorphic quadratic differential with only simple poles at the vertex operator insertion points:
\begin{multline}\label{gP2}
P^2(z)=\sum_{I,J=1}^{g}\ell^{I}\cdot\ell^{J} \omega_{I}(z)\omega_{J}(z)+2\sum_{I=1}^{g}\sum_{i=1}^{n}\ell^{I}\cdot k_{i}\, \omega_{I}(z) \tilde{S}_{g}(z,z_i|\Omega) \\
+\sum_{i\neq j}k_{i}\cdot k_{j}\, \tilde{S}_{g}(z,z_i|\Omega)\tilde{S}_{g}(z,z_j|\Omega)\,.
\end{multline}
The vectors $\{\ell_{m}^{I}\}$ are the zero modes of $P_{m}$, associated with homogeneous solutions of \eqref{Peom}, whereas the residue of $P^2$ at $z_i$ is easily seen to be
\be\label{Pres}
 \mathrm{Res}_{z=z_i}P^2(z)=\sum_{I=1}^{g}k_{i}\cdot\ell^{I}\,\omega_{I}(z_i)+\sum_{j\neq i}k_{i}\cdot k_{j}\,\tilde{S}_{g}(z_i,z_j|\Omega)\,.
\ee

In light of \eqref{gP2}, the delta functions appearing in the correlators \eqref{ampg}, \eqref{amp1}, \eqref{amp0} have a natural interpretation: they enforce the condition that $P^2=0$ globally on the worldsheet $\Sigma$. As noted in~\cite{Adamo:2013tsa}, this is the geometric content of the \emph{scattering equations} at generic genus. Indeed the amplitude prescription ensures that there are $3g-3+n$ delta function constraints for $g\geq 2$: $n$ of them to set the residues \eqref{Pres} to zero, and $3g-3$ to kill the remaining globally-defined moduli. At $g=0,1$ this counting is modified in the obvious way in accordance with $h^{0}(\Sigma, K^{2}_{\Sigma}(z_{1}+\cdots+z_n))$.

Hence, it is clear that the amplitude prescription will give the expected scattering equations at a given genus, along with a non-compact zero-mode integral over the $\{\ell^{I}_{m}\}$. These scattering equations completely fix all the moduli integrals (over $\{\tau_{a}\}$ and $\{z_{i}\}$) in terms of the kinematics (the external and loop momenta $\{k_{i},\ell^{I}\}$). We therefore deduce that a general amplitude will take the form:
\begin{eqnarray}
 \cM_{n}^{(g)} & = & \delta^{10}\left(\sum_{i=1}^{n}k_i^{m}\right)\int \prod_{I=1}^{g}\d^{10}\ell^{I}\,\prod_{a=1}^{3g-3}\d \tau_{a} \bar{\delta}\left(P^2(z_a)\right) \prod_{j=1}^{n}\bar{\delta}\left(k_{j}\cdot P(z_j)\right) \left\la \cN \widetilde{\cN}\,\cdots \right\ra \notag \\
 & := & \delta^{10}\left(\sum_{i=1}^{n}k_i^{m}\right)\int \prod_{I=1}^{g}\d^{10}\ell^{I}\, \mathfrak{M}^{(g)}_{n}\,,
\end{eqnarray}
where the integrand $\mathfrak{M}^{(g)}_{n}$ represents the full correlator, localized on the support of the scattering equations with all OPEs and zero mode integrations performed, except for the loop integrals $\d^{10}\ell$. 

Our central conjecture is that the quantity $\mathfrak{M}^{(g)}_{n}$ is equal to the $g$-loop \emph{integrand} of type II supergravity, before any loop integrals have been performed. By the `integrand', we mean the sum over all $g$-loop Feynman diagrams in the field theory without performing the loop integrations. Although type II supergravity is UV divergent in ten-dimensions, we expect these divergences to emerge only after performing the $\d^{10}\ell$ integrals, so the integrand $\mathfrak{M}^{(g)}_{n}$ itself is a perfectly well-defined object. 

Of course, it is far from obvious that the worldsheet correlators will have even the most rudimentary properties of field theory amplitudes, such as being rational functions of the kinematic data, producing the correct kinematic prefactors, or factorizing correctly. However, we will show that in the special case of the four-point amplitudes, the correlators do indeed pass several non-trivial tests in favor of the conjecture. In particular, we recover the correct kinematic prefactor and IR behavior consistent with supergravity amplitudes. These tests are enabled by a combination of similar results for the higher-genus amplitudes of the RNS-like formalism~\cite{Adamo:2013tsa,Casali:2014hfa}, as well as the similarities between this worldsheet theory and the non-minimal formalism of the pure spinor superstring, where extensive calculations have been performed explicitly. 

\medskip

At genus zero, there are no zero modes of $P_{m}$ to integrate over and the conjecture reduces to the claim that $\cM^{(0)}_{n}$ gives the full tree-level S-matrix of type II supergravity. On the genus zero worldsheet, the regulator is simply
\begin{equation*}
 \cN=\e^{-\lambda\cdot\bar{\lambda}-r\cdot\theta}\,,
\end{equation*}
since none of the conformal weight $(1,0)$-fields have any zero modes. Performing the $X$ path integral fixes $P_{m}$ via \eqref{gP} to be
\begin{equation}
 P_{m}(z)=\d z\,\sum_{i=1}^{n}\frac{k_{i\;m}}{z-z_{i}}\,, 
\end{equation}
so all the remaining fields in the correlator \eqref{amp0} are the \emph{same} as left-moving variables of the superstring. After contracting all the conformal weight $(1,0)$ fields via their OPEs, the same strategy as the superstring~\cite{Berkovits:2005bt} reveals that after integrating out the non-minimal variables,
\begin{equation}
 \cM_{n}^{(0)}=\int [\d\lambda][\d\tilde{\lambda}]\,\d^{16}\theta\, \d^{16}\tilde{\theta}\, \prod_{i=4}^{n}\bar{\delta}\left(k_i\cdot P(z_i)\right) \lambda^{\alpha}\lambda^{\beta}\lambda^{\gamma}\til{\lambda}^{\til\alpha}\til\lambda^{\til\beta}\til\lambda^{\til\gamma} f_{\alpha\til{\alpha}\beta\til{\beta}\gamma\til{\gamma}}(\theta,\til{\theta}) \,,
\end{equation}
where $f$ is a function of the kinematic data, the insertion points, and takes values in $\otimes_{i=4}^{n} K^{2}_{\Sigma \;i}$.

It is easy to see that this is equivalent to the minimal prescription \eqref{minamp1} given by Berkovits~\cite{Berkovits:2013xba}, and in turn proven to give the full tree-level S-matrix of supergravity~\cite{Gomez:2013wza}. So at genus zero, the non-minimal formalism reduces to the minimal formalism in exactly the same way as for superstring theory, and gives the desired classical scattering amplitudes of type II supergravity.\footnote{In principle, one could define a higher genus prescription for the minimal model analogous to the superstring. While avoiding this for the reasons mentioned above, we expect an abstract equivalence between the two formalisms to hold beyond tree-level, again in analogy with superstring theory~\cite{Hoogeveen:2007tu}.}  We now turn to some explicit calculations of the four-point function to provide evidence for the conjecture beyond tree-level.


\subsection{Four-point function: Genus one}

On a genus one surface the fields of conformal weight $(1,0)$ acquire zero modes. In particular the fermionic fields $s^\alpha$ and $\tilde{s}^{\til\alpha}$ have 11 zero modes each, which must be soaked up by operator insertions in the path-integral to give a non-vanishing result. The only operators which can provide these zero modes are the regulators $\mathcal{N}$ and $\widetilde{\mathcal{N}}$, given at genus one by~\cite{Berkovits:2005bt}
\begin{align}
 \mathcal{N}=\exp\left(-\lambda\cdot\bar\lambda-r\cdot\theta-w_{\mathrm{z.m.}}\cdot\bar{w}_{\mathrm{z.m.}}+s_{\mathrm{z.m.}}\cdot d_{\mathrm{z.m.}}\right)\,,
\end{align}
where $f_{\mathrm{z.m.}}$ denotes the zero mode of the conformal weight $(1,0)$ field $f$. The 11 zero modes of $s^{\alpha}$ are thus accompanied by 11 zero modes of the $d_{\alpha}$ field, the latter of which has 16 unconstrained components. So there are 5 remaining zero modes of $d_{\alpha}$ left to be soaked up by contributions coming either from vertex operators or the $b$-ghost insertion in \eqref{amp1}. 

Fixed vertex operators cannot contribute $d$ zero modes, so they must come either from integrated vertex operators, which can contribute at most one $d$ zero mode each, or from the effective $b$-ghost, which can contribute at most 2 zero modes. The counting is exactly the same for the tilded variables. Using this zero mode counting, it is clear that the first non-vanishing amplitude at genus one is the four-point amplitude; $\cM^{(1)}_{n<4}=0$ since the fermionic zero mode integrals cannot be saturated. This vanishing is a consequence of spacetime supersymmetry, which is manifest in the pure spinor approach. In the RNS-like model, the vanishing of the lower point amplitudes occurs only after summing over spin structures~\cite{Adamo:2013tsa}.

At four points there is only one way to pick terms from the vertex operators and $b$-ghost in order to saturate the $d$ zero mode path integral, just as in superstring theory~\cite{Berkovits:2004px,Mafra:2005jh,Berkovits:2006bk}. Each of the three integrated vertex operators \eqref{iVO} contributes a zero mode from the term $d_\alpha W^\alpha$ and the $b$-ghost \eqref{effbg} contributes 
\begin{equation*}
(b|\mu)\propto \frac{(\bar\lambda\gamma_{mnp}r)(d_{\mathrm{z.m.}}\gamma^{mnp}d_{\mathrm{z.m.}})}{(\bar\lambda\cdot\lambda)^2}\,.
\end{equation*}
After performing the $d$ zero mode integral we are left with
\begin{align}
\int \d^{16}\theta\,\int[\d\lambda][\d\bar\lambda][\d r]\,\frac{(\bar{\lambda}\gamma_{mnp}D)}{(\bar\lambda\cdot\lambda)^2}(\lambda\cdot A_{1})(\lambda\gamma^{m}W_{2}) (\lambda\gamma^{n}W_{3}) (\lambda\gamma^{p}W_{4})\,\e^{-\lambda\cdot\bar{\lambda}-r\cdot\theta}\,.
\end{align}
This has been shown \cite{Berkovits:2006bk,Mafra:2009wq} to be proportional to the pure spinor superspace expression
\begin{align}\label{1kpf}
K = \left\la (\lambda\cdot A_{1})(\lambda\gamma_{m} W_{2})(\lambda\gamma_n W_{3})\mathcal{F}^{mn}_{4}\right\ra\,,
\end{align}
where these angle brackets stand for the pure spinor and theta zero mode integrations. The calculation in the tilded variables is identical. Thus, the amplitude can be written as
\begin{align}\label{4pg1}
 \mathcal{M}^{(1)}_4\propto K\,\widetilde{K}\,\int \d^{10}\ell\int \d\tau\,(\d z_0)^2\,\bar\delta\left(P^2(z_0)\right)\,\prod_{i=2}^{4} \bar\delta(k_i\cdot P(z_i))\,(\d z_i)^2\,,
\end{align}
omitting the overall momentum conserving delta function. 

The amplitude \eqref{4pg1} is equal to the amplitude given by the RNS-like formalism after summing over spin structures~\cite{Adamo:2013tsa}, with $K\widetilde{K}$ the correct supersymmetric prefactor for supergravity. As expected, the integral over the moduli space of a four-punctured torus is completely localized by the scattering equations.\footnote{The integrand is seen to be modular invariant by adopting the prescription of~\cite{Adamo:2013tsa} to demand that $P_{m}$ transforms invariantly.} It is straightforward to see that this integrand factorizes with a simple pole in the $\tau\rightarrow i\infty$ limit onto a rational function of the kinematics, and in the IR region (where three adjacent propagators can be thought of as going on-shell) it reproduces the field theoretic sum over scalar box integrals~\cite{Casali:2014hfa}. Hence, the four-point function factorizes as a field theory amplitude, and is explicitly equal to the supergravity amplitude deep in the IR.


\subsection{Four-point function: Genus two}

By now it should be clear that computations involving only zero mode counting in this model will be almost the same as in the usual pure spinor superstring. In particular, the computation of the genus two four-point amplitude can be carried out in much the same way as in the pure spinor superstring. In this case there are now 22 zero modes of the field $s$; these, again, can only come from the regulators and thus are accompanied by 22 zero modes of the $d$ field. At genus two the field $d$ has 32 zero modes, so 10 other zero modes must be provided by the integrated vertex operators and $b$-ghosts. Each integrated vertex operator can contribute at most one zero mode, so we need to take two zero modes from each $b$-ghost. 

This completely fixes the terms we take from each operator, which are the same as in the one-loop case. After doing the path integral over the zero modes of $d,\;s,\;w,$ and $\bar w$, the remaining superspace expression can be written as~\cite{Berkovits:2005df,Berkovits:2006bk,Gomez:2010ad,Mafra:2009wq}
\begin{equation*}
 \int \d^{16}\theta\,\int[\d\lambda][\d\bar\lambda][\d r]\,\frac{(\lambda\gamma_{mnpqr}\lambda)}{(\bar{\lambda}\cdot\lambda)^3} \mathcal{F}^{mn}\mathcal{F}^{pq}\mathcal{F}^{rs}(\lambda\gamma_{s}W)\,\e^{-\lambda\cdot\bar{\lambda}-r\cdot\theta}\,,
\end{equation*}
where we suppress various numerical factors and the distribution of particle labels on the superfields. Upon summing over permutations of particle labels, this superspace expression vanishes unless it is dressed with holomorphic differentials arising from a combination of the moduli integrals and the $b$-ghost insertions. The result can be identified with the kinematic prefactor of supergravity~\cite{Berkovits:2005df,Berkovits:2005ng} by comparison with the computation in the RNS formalism~\cite{D'Hoker:2001nj,D'Hoker:2005jc}, or via BRST cohomology arguments~\cite{Mafra:2008ar}. The counting and calculation for the tilded variables follows identically. 

The BRST cohomology techniques of~\cite{Mafra:2008ar} relate the two-loop kinematic prefactor to the one-loop prefactor \eqref{1kpf}, see also~\cite{Gomez:2010ad}. Applying this relationship leaves us with the expression
\be\label{two-loop}
 \cM^{(2)}_{4}\propto K\,\widetilde{K}\,\int \d^{10}\ell_{1}\,\d^{10}\ell_{2} \int \d^{3}\Omega\;\mathcal{Y}^2 \prod_{j=1}^{3}\bar{\delta}\left(P^2(x_j)\right)\,(\d x_j)^2 \, \prod_{i=1}^{4}\bar{\delta}\left(k_i\cdot P(z_i)\right) \,,
\ee
where $\d^3\Omega$ stands for the integrals over the complex structure moduli of the genus two Riemann surface and $\mathcal{Y}$ is the quadri-holomorphic form~\cite{D'Hoker:2005jc}
\begin{align}\label{qform}
 \mathcal{Y}=(t-u)\Delta(1,2)\Delta(3,4)+(s-t)\Delta(1,3)\Delta(4,2)+(u-s)\Delta(1,4)\Delta(2,3)\,.
\end{align}
Here, $\{s,t,u\}$ are the standard Mandelstam parameters (\textit{e.g.}, $s=2k_{1}\cdot k_{2}$) and 
\begin{equation*}
 \Delta(z,w)=\omega_1(z)\omega_2(w)-\omega_1(w)\omega_2(z)
\end{equation*}
for $\omega_I$ the abelian differentials on the genus two worldsheet.

\medskip

\begin{figure}[t]
\centering
\includegraphics[width=2.0 in, height=1.0 in]{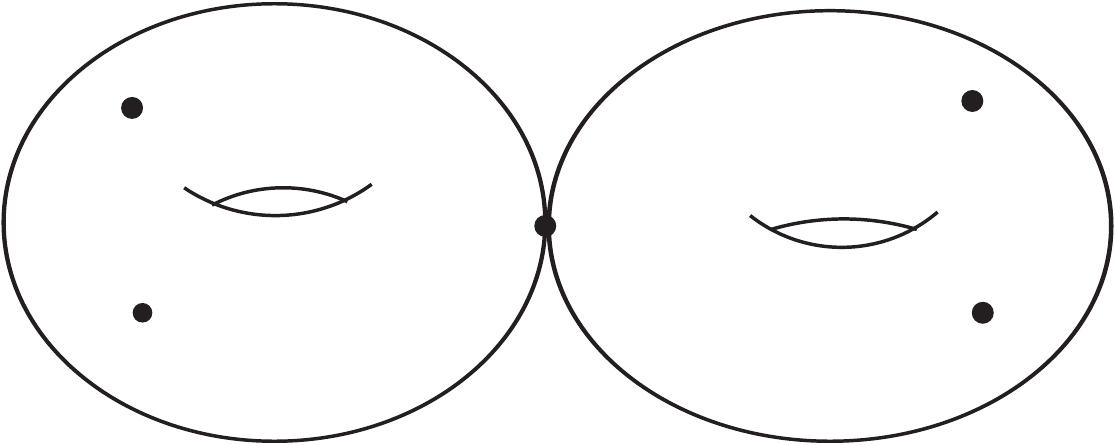}\caption{\small{\textit{The genus two worldsheet at the boundary of the moduli space.}}}\label{2lf2}
\end{figure}

Our conjecture is that the integrand of \eqref{two-loop} is a representation for the two-loop integrand of type IIA/B supergravity. In particular, the massive modes that usually run through the loops of string theoretic amplitudes at genus two should be absent. There is an easy test that can be done in this amplitude to show that no massive modes are propagating. We can look at the boundary of the moduli space where the genus two surface degenerates into two tori glued at a nodal point, see Figure~\ref{2lf2}. In the superstring the only poles at this boundary come from the propagation of massive modes through the node~\cite{D'Hoker:2005ht}. In terms of the field theory integrand, this boundary corresponds to a non-existent cut of a double box. Therefore if \eqref{two-loop} represents a field theory amplitude, it must vanish at this separating boundary.

Using the period matrix
\begin{equation*}
 \Omega=\left(\begin{array}{cc}
               \tau_{11} & \tau_{12}\\
               \tau_{12} & \tau_{22}
              \end{array}\right)
\end{equation*}
to parametrize the genus two surface, the separating boundary divisor of the moduli space sits at $\tau_{12}\rightarrow0$; $\tau_{11},\tau_{22}$ are the modular parameters of the two resulting tori. Near this boundary we let the external states $\{1, 2\}$ move to one of the tori, call it $\Sigma_1$ with modular parameter $\tau_{11}$, while states $\{3,4\}$ move to the other torus, call it $\Sigma_2$ with modular parameter $\tau_{22}$. With this choice, the quadri-holomorphic form \eqref{qform} becomes simply~\cite{D'Hoker:2005ht}
\begin{align}\label{quadlim}
 \mathcal{Y}\xrightarrow{\tau_{12}\rightarrow0} -s=-2k_{1}\cdot k_{2}\,,
\end{align}
with no pole arising from the measure factors. We expect that at this boundary, the scattering equations in \eqref{two-loop} enforce the momentum flowing through the node to be on-shell (\textit{i.e.}, $s=0$) and thus the amplitude vanishes. 

At this stage, it is convenient to make use of an explicit parametrization of the moduli space near this boundary, which has been deployed often in the study of factorization in string theory (the so-called `plumbing fixture' \textit{c.f.}, \cite{Fay:1973,Verlinde:1986kw,Polchinski:1988jq,D'Hoker:1988ta}). On the two tori $\Sigma_1, \Sigma_{2}$ pick local coordinates $z_I$ around one point on each surface $p_I\in\Sigma_I$ such that $p_I=\{z_I=0\}$. Remove an open neighborhood around these points $U_I=\{|z_I|<|t|^{1/2}\}$ where $t$ is a coordinate on the unit disk $D=\{t\in\mathbb{C}|\;|t|<1\}$ (not to be confused with the Mandelstam variable, which we no longer need). Now glue them together using the annulus $A_t=\{w\in\mathbb{C}|\;|t|^{1/2}<|w|<|t|^{-1/2}\}$ via 
\begin{align}
 w = \begin{cases}\displaystyle
      \frac{t^{1/2}}{z_1}\;&\text{if}\;\; |t|^{1/2}<|w|<1\\
      t^{-1/2}\, z_{2}\;&\text{if}\;\; 1<|w|<|t|^{1/2}
     \end{cases}\,.
\end{align}

\begin{figure}[t]
\centering
\includegraphics[width=3.4 in, height=1.4 in]{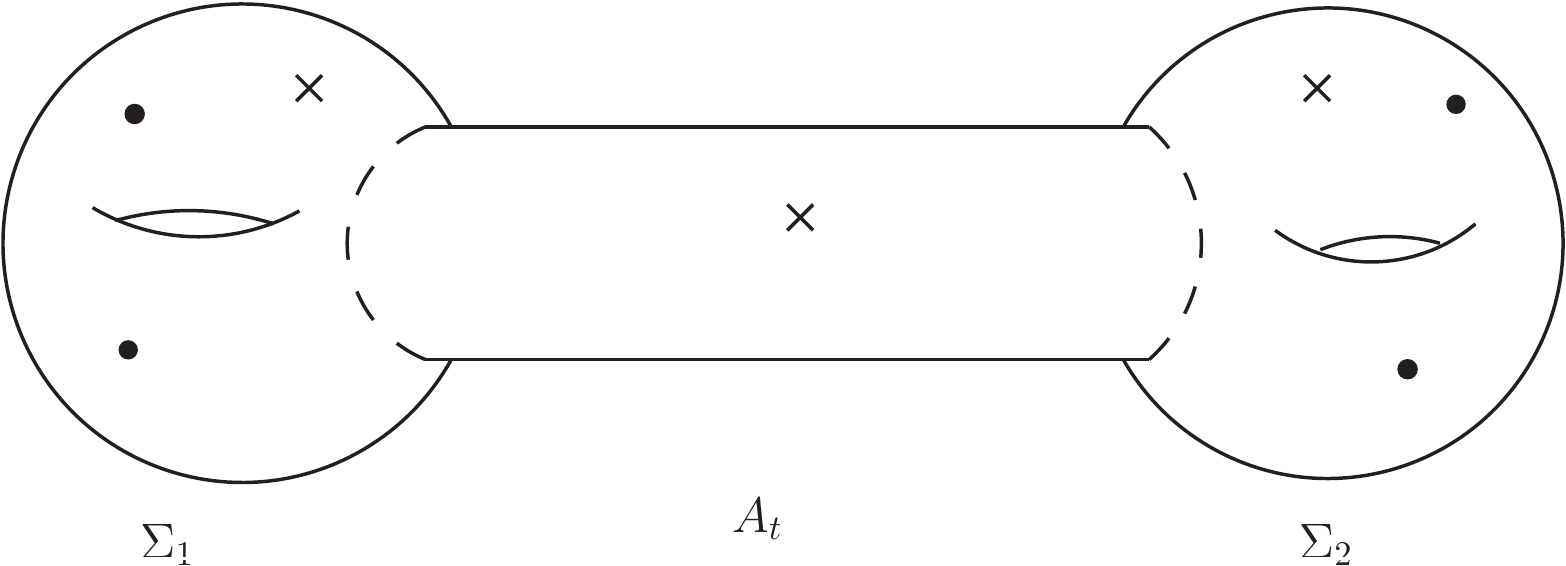}\caption{\small{\textit{The degenerating worldsheet modeled on two tori $\Sigma_1, \Sigma_2$ connected by the annulus $A_t$. Solid dots denote scattering equations of the form $k\cdot P$, while crosses denote scattering equations of the form $P^2$. }}}\label{2lf1}
\end{figure}

This gives a family of genus two Riemann surfaces fibered over the unit disk which can be seen as the union of three distinct components, $(\Sigma_1\setminus U_1)\cup A_t\cup(\Sigma_2\setminus U_2)$. The singular fiber over $t=0$ corresponds to the boundary we are interested in, and one can show that $t\propto\tau_{12}$. We now distribute the scattering equations among these components. The four scattering equations of the form $k_{i}\cdot P(z_i)$ accompany the punctures, so the $i=1,2$ equations go to $\Sigma_1\setminus U_1$, while $i=3,4$ go to $\Sigma_2\setminus U_2$.  There are also three $P^2(x)$ scattering equations, corresponding to each of the three moduli of the genus two surface. The natural choice is to place one of these equations on each component of the family of surfaces (see Figure~\ref{2lf1}). 

The form of these equations as we approach the boundary is dictated by the field $P_{m}(z)$, whose behavior under the degeneration depends on which component it is being evaluated at. Using standard degeneration formulas for the abelian differentials and propagators, it is easy to see what happens to $P$. The abelian differentials behave as~\cite{Fay:1973}
\begin{align}\label{diffgen}
 \omega_{I}(z)=\begin{cases}
             \varpi_{I}(z) +O(t) &\;\text{if}\; z\in\Sigma_I\\
             O(t) &\text{otherwise }
            \end{cases}\,, \qquad I=1,2\,,
\end{align}
where $\varpi_I$ are the global holomorphic differentials on the respective tori. The behavior of the propagator $\tilde{S}_{2}(z,w|\Omega)$ can be deduced from that of the prime form
\begin{align}\label{primegen}
 E_2(z,w|\Omega)=\begin{cases}
                  -E_1(z,p_1|\tau_{11})wt^{-1/4} &\text{if } z\in\Sigma_1,\;w\in A_t\\
                  E_1(z,p_2|\tau_{22})t^{-1/4} &\text{if } z\in\Sigma_2,\;w\in A_t\\
                  E_1(z,p_1|\tau_{11})E_1(p_2,w|\tau_{22})t^{-1/2} &\text{if } z\in\Sigma_1,\;w\in\Sigma_2
                 \end{cases}\,.
\end{align}

Using \eqref{diffgen}--\eqref{primegen} with \eqref{gP} it is straightforward to see that as $t\rightarrow 0$, the scattering equations on each component $\Sigma_I\setminus U_I$ go to the one-loop scattering equations with an extra puncture at $p_I$ of momentum $\pm(k_1+k_2)$. This is a consequence of
\begin{equation*}
 \lim_{t\rightarrow 0}P_{m}(z)|_{\Sigma_1}=\ell_{1\;m}\,\varpi_{1}(z)+\sum_{i=1,2}k_{i\;m}\tilde{S}_{1}(z,z_i|\tau_{11})-(k_1+k_2)_{m} \tilde{S}_{1}(z,p_1 |\tau_{11})\,,
\end{equation*}
\begin{equation*}
 \lim_{t\rightarrow 0}P_{m}(z)|_{\Sigma_2}=\ell_{2\;m}\,\varpi_{2}(z)+\sum_{i=3,4}k_{i\;m}\tilde{S}_{1}(z,z_i|\tau_{22})+(k_1+k_2)_{m} \tilde{S}_{1}(z,p_2 |\tau_{22})\,.
\end{equation*}
The remaining scattering equation on the annulus enforces the momentum flowing through the node to be on-shell, since
\begin{align}
 P_{m}(w)= (k_{1}+k_{2})_{m}\,\frac{\d w}{w} + O(t) \:\:\:\mbox{if } w\in A_t\,,
\end{align}
where $\frac{\d w}{w}$ is the holomorphic differential on the annulus. 

Therefore $P^2(w)\propto s+O(t)=0$, enforcing $k_{1}\cdot k_{2}=0$ in the $t\rightarrow0$ limit. Since the amplitude in this limit is multiplied by a factor of $s$ from \eqref{quadlim}, it vanishes on top of the scattering equations. This gives further evidence that the model describes only field theory amplitudes in type IIA/B supergravity. For more general external kinematics, it should also be possible to extract scalar integrals from \eqref{two-loop} by probing the deep IR behavior of the integrand (\textit{i.e.}, considering multiple adjacent propagators going on-shell) and using the techniques of~\cite{Casali:2014hfa}. The squared Mandelstam invariants, which accompany each planar or non-planar double box in four-point two-loop supergravity amplitudes~\cite{Bern:1998ug}, are supplied by $\mathcal{Y}^2$.

\medskip

Note that although we confined our attention to four-point amplitudes in this section, the factorization arguments for the scattering equations generalize in the obvious way to $n$-points and arbitrary genus. Combined with the non-separating degenerations (\textit{i.e.}, pinching a cycle of the worldsheet non-homologous to zero) and the separating degenerations that pinch off a sphere from the worldsheet, both of which were studied in~\cite{Adamo:2013tsa} making use of methods from~\cite{Adamo:2013tca}, this encompasses \emph{all} possible degenerations of the scattering equations near a boundary of the moduli space for any genus and any number of external states.


\section{Conclusions}
\label{sec:concl}

In this paper, we showed how to define a non-minimal formalism for the worldsheet theory which appears to describe the S-matrix of type II supergravity in ten dimensions. Following the analogy with superstring theory, we arrived at an amplitude prescription on arbitrary genus worldsheets, which we conjectured to provide the \emph{loop integrands} of supergravity, with UV divergences emerging from the non-compact loop integrations $\prod_I \d^{10}\ell_I$. By performing some explicit four-point calculations at genus one and two, we found non-trivial evidence in favor of this conjecture: the worldsheet correlators produce the kinematic prefactors of the superstring and have factorization behavior in accordance with field theory amplitudes.

Besides the obvious question as to whether the conjecture can be proven (in the pure spinor context or the RNS setting of~\cite{Adamo:2013tsa}), this work also hints towards intriguing relationships with other results in the literature. While the worldsheet model presented here or in~\cite{Berkovits:2013xba} can be thought of heuristically as an `infinite tension limit' of the superstring, it is more precisely related to a holomorphic complexification of the pure spinor worldline formalism~\cite{Berkovits:2001rb}. At tree-level and for the minimal formalism, this is quite clear: the fixed vertex operators are identical, and integrated vertex operators are complexified at the expense of a delta function factor which leads to the scattering equations.

The non-minimal formalism is likewise related to a non-minimal worldline action presented in~\cite{Bjornsson:2010wu,Bjornsson:2010wm}. However, the similarities extend beyond the worldsheet/line actions: the loop amplitude prescription for the worldsheet is easily seen to be a sort of complexification of the rules developed for the worldline. The efficacy of the worldline formalism in studying the UV divergences of (dimensionally-reduced) maximal supergravity would seem to provide additional -- admittedly heuristic -- evidence in favor of our conjecture for the worldsheet model. Finding a more concrete connection between the worldline formulation of supergravity and the worldsheet model could lead to a quantitative proof of these ideas.

Along these lines, it is interesting to note a sort of `conservation of difficulty' in evaluating amplitude expressions, even at four points. In the worldsheet model, the correlator is easy to compute, boiling down to nothing more than zero mode counting and, for genus greater than two, non-trivial OPEs between worldsheet fields (as in the superstring~\cite{Gomez:2013sla}). However, the integrand is then localized on the support of the genus $g$ scattering equations, which fix the moduli in terms of the kinematic data and loop momenta via elliptic functions. Consequently, solving these equations analytically or numerically for general kinematics seems prohibitively difficult. By contrast, there are no scattering equations to solve in the worldline formalism, but a sum over skeleton graphs with increasingly varied topology for $g>2$ must be performed explicitly. It seems that one trades this sum over graphs for localization onto the scattering equations by moving from the worldline to worldsheet descriptions of perturbative supergravity.

\medskip

A similar phenomenon also exists upon comparison with another prescription which has been developed for the computation of field theory loop integrands based on the pure spinor formalism~\cite{Mafra:2010ir,Mafra:2014oia,Mafra:2014gja,Mafra:2015gia}. This method exploits the underlying superstring origin of field theory amplitudes to construct BRST-invariant integrands using multiparticle superfields derived from OPE computations in the pure spinor formalism. The expected gauge anomaly for higher-point amplitudes is accounted for by pseudo-BRST invariants constructed in~\cite{Mafra:2014gsa}. These multiparticle building blocks are then fused together to build amplitudes from trivalent graphs. This program has now been explicitly realized at tree-level (for the amplitude)~\cite{Mafra:2010ir,Mafra:2010jq} and one-loop (for the integrand)~\cite{Mafra:2014gja}, and aims to provide a mapping between trivalent graphs and multiparticle superfields at \emph{any} level in perturbation theory.

If our conjecture is true, then the integrands $\mathfrak{M}^{(g)}_{n}$ computed by the worldsheet model should be equal to those arising from the BRST cohomology approach. However, this would entail trading localization onto the support of the scattering equations (a process which seems very difficult beyond tree-level and a low number of external states) for the sum over trivalent graphs expressed in terms of multiparticle superfields (which produces analytic and manifestly supersymmetric answers). At tree-level, the translation to trivalent graphs is implied by the scattering equations and the correlator structure~\cite{Cachazo:2013iea,Gomez:2013wza}; any understanding of how this relationship could be realized at higher loops would be very interesting.

\medskip

Finally, we note that a heterotic version of the minimal model was also presented in~\cite{Berkovits:2013xba}, with vertex operators coupled to the worldsheet current algebra giving the superfields of $\cN=1$ super-Yang-Mills in ten-dimensions. The sphere correlation functions of these vertex operators were shown to give the correct tree-level S-matrix for SYM in $d=10$. However, this model also contains gravitational vertex operators, which do \emph{not} produce the scattering amplitudes of $\cN=1$ supergravity. Hence, higher-genus amplitudes in the heterotic model will be contaminated by these unphysical states running in the loop. A similar issue exists for the RNS-like formulation of this heterotic model~\cite{Mason:2013sva} where the problem can also be understood from the absence of an anomaly cancellation mechanism under target space diffeomorphisms~\cite{Adamo:2014wea}. Insight into how to formulate these worldsheet theories for Yang-Mills degrees of freedom alone or coupled to Einstein (super)gravity would enable the study of higher-genus worldsheet descriptions of gauge theory amplitudes as well. 

\acknowledgments

We thank Carlos Mafra, Oliver Schlotterer, and Piotr Tourkine for many helpful discussions and comments on the manuscript, as well as Michael Green and David Skinner for useful conversations. The work of TA is supported by a Title A Research Fellowship at St. John's College, Cambridge. The work of EC is supported in part by the Cambridge Commonwealth, European and International Trust.


\appendix

\section{Worldsheet currents and OPEs}
\label{app:currents}

For convenience, we list here the various composite currents appearing in the non-minimal formalism discussed in this paper, and their various OPEs with each other. We only list these for the un-tilded variables, as the currents and OPEs for the tilded sector are identical. Note that the only difference between the list here and that for the superstring is the definition of the Green-Schwarz current $d_{\alpha}$ (\textit{c.f.}, \cite{Bedoya:2009np}).

The pure spinor conditions on minimal and non-minimal variables imply a gauge invariance, meaning that the conformal weight $(1,0)$ pure spinor fields can only appear in the currents:
\begin{equation*}
 N^{nm}=\frac{1}{2}(w\gamma^{nm}\lambda)\,, \qquad J=\lambda\cdot w\,, \qquad T_{\lambda}=-w_{\alpha}\,\partial \lambda^{\alpha}\,,
\end{equation*}
\begin{equation*}
 \bar{N}^{nm}=\frac{1}{2}\left(\bar{w}\gamma^{nm}\bar{\lambda}+s\gamma^{nm}r\right)\,, \qquad \bar{J}=\bar{w}\cdot\bar{\lambda}+s\cdot r\,, \qquad T_{\bar{\lambda},r}=-\bar{w}^{\alpha}\partial\bar{\lambda}_{\alpha}-s^{\alpha}\partial r_{\alpha}\,,
\end{equation*}
\begin{equation*}
S_{mn}=\frac{1}{2}(s\gamma_{mn}\bar{\lambda})\,, \qquad S=s\cdot\bar{\lambda}\,.
\end{equation*}
The minimal currents have OPEs:
\begin{equation*}
 N^{nm}(z)\,\lambda^{\alpha}(w)\sim -\frac{1}{2}\frac{(\gamma^{nm}\lambda)^{\alpha}}{z-w}\,, \qquad J(z)\,\lambda^{\alpha}(w)\sim -\frac{\lambda^{\alpha}}{z-w}\,, \qquad J(z)\,N^{nm}(w)\sim 0 \,,
\end{equation*}
\begin{equation*}
 J(z)\,J(w)\sim \frac{-4}{(z-w)^2}\,, \qquad N^{pq}(z)\,N^{nm}(w)\sim -3\frac{\eta^{m[p}\eta^{q]n}}{(z-w)^2}+\frac{\eta^{m[q}N^{p]n}-\eta^{n[q}N^{p]m}}{z-w}\,,
\end{equation*}
\begin{equation*}
 N^{nm}(z)\,T_{\lambda}(w)\sim \frac{N^{nm}(z)}{(z-w)^2}\,, \qquad J(z)\,T_{\lambda}(w)\sim \frac{8}{(z-w)^3}+\frac{J(z)}{(z-w)^2}\,,
\end{equation*}
\begin{equation*}
 T_{\lambda}(z)\,T_{\lambda}(w)\sim \frac{11}{(z-w)^4}+\frac{2\,T_{\lambda}(z)}{(z-w)^2}+\frac{\partial T_{\lambda}}{z-w}\,.
\end{equation*}
The last of these confirms the $+22$ central charge contribution of the pure spinor variables.

For the non-minimal variables, we have:
\begin{equation*}
 \bar{N}^{nm}(z)\,\bar{\lambda}_{\alpha}(w)\sim -\frac{1}{2}\frac{(\gamma^{nm}\bar{\lambda})_{\alpha}}{z-w}\,, \qquad \bar{N}^{nm}(z)\,r_{\alpha}(w)\sim -\frac{1}{2}\frac{(\gamma^{nm}r)_{\alpha}}{z-w}\,, \qquad \bar{J}(z)\,\bar{N}^{nm}(w)\sim 0\,,
\end{equation*}
\begin{equation*} 
 \bar{J}(z)\,\bar{\lambda}_{\alpha}(w)\sim -\frac{\bar{\lambda}_{\alpha}}{z-w}\,, \qquad \bar{J}(z)\,r_{\alpha}(w)\sim -\frac{r_{\alpha}}{z-w}\,, \qquad \bar{J}(z)\,\bar{J}(w)\sim 0\,,
\end{equation*}
\begin{equation*}
 \bar{N}^{pq}(z)\,\bar{N}^{nm}(w)\sim \frac{\eta^{m[q}\bar{N}^{p]n}-\eta^{n[q}\bar{N}^{p]m}}{z-w}\,,
\end{equation*}
\begin{equation*}
 \bar{N}^{nm}(z)\,T_{\bar{\lambda},r}(w)\sim \frac{\bar{N}^{nm}(z)}{(z-w)^2}\,, \qquad \bar{J}(z)\,T_{\bar{\lambda},r}(w)\sim \frac{\bar{J}(z)}{(z-w)^2}\,,
\end{equation*}
\begin{equation*}
 T_{\bar{\lambda},r}(z)\,T_{\bar{\lambda},r}(w)\sim \frac{2\,T_{\bar{\lambda},r}(z)}{(z-w)^2}+\frac{\partial T_{\bar{\lambda},r}}{z-w}\,,
\end{equation*}
Any additional OPEs (involving $S_{nm}$ or $S$) can be read off directly from the superstring (\textit{c.f.}, \cite{Berkovits:2005bt}). Note that the OPE of the stress tensor $T_{\bar{\lambda},r}$ with itself confirms that the non-minimal variables do not modify the central charge of the model.

Finally, the BRST charge for both the minimal and non-minimal models is built upon the Green-Schwarz constraint $d_{\alpha}$ \eqref{GScon}, which is the holomorphic generalization of the superparticle constraint:
\begin{equation*}
 d_{\alpha}=p_{\alpha}-\frac{1}{2}P_{m}\gamma^{m}_{\alpha\beta}\theta^{\beta}.
\end{equation*}
The OPEs of this constraint with the other matter variables are crucial for proving nilpotence of the BRST charge, closure of the vertex operators, as well as deriving the effective $b$-ghost \eqref{effbg}, and can be derived using the free OPEs \eqref{matOPE}:
\begin{equation*}
 d_{\alpha}(z)\,f\left(X,\theta\right)(w)\sim \frac{D_{\alpha} f}{z-w}\,, \qquad d_{\alpha}(z)\,d_{\beta}(w)\sim -\frac{P_{m}\gamma^{m}_{\alpha\beta}}{z-w}\,,
\end{equation*}
where $D_{\alpha}$ is the supersymmetric derivative.


\section{Worldsheet action from a gauge-fixing procedure}
\label{app:gf}

The origins of the pure spinor formalism for the superstring are still somewhat shrouded in mystery, particularly with regards to the emergence of the worldsheet action from a gauge fixing procedure, as in the RNS formalism for superstring perturbation theory. Recently, Berkovits showed that the pure spinor superstring can be derived from gauge fixing a reparametrization-invariant action~\cite{Berkovits:2014aia}. The starting point is a worldsheet theory with only bosonic variables; one then gauges a twistor-like constraint (along with the ten-dimensional pure spinor $\lambda^{\alpha}$) to arrive at the pure spinor formalism. Remarkably, all fermionic worldsheet variables emerge as ghosts in this gauge fixing procedure. While this process still requires the `by hand' imposition of the pure spinor condition, it does give a derivation of the superstring action from a first principles argument and also hints at a surprisingly universal role for twistor-like geometry in superstring theory.

The worldsheet model studied in this paper can also be derived by a very similar procedure. This also involves gauging a twistor-like constraint on the worldsheet, but in this context the constraint seems to be stronger, implying both the Virasoro constraint and the Hamiltonian. We expect this fact to be related to some of the curiosities appearing in our amplitude prescription; for instance, that the effective $b$-ghost gives a prescription for performing the moduli integrals, while being related to the Hamiltonian rather than the stress tensor.

We begin with a chiral, first-order version of the reparametrization-invariant worldsheet action from~\cite{Berkovits:2014aia}:
\be\label{gf1}
S=\frac{1}{2\pi}\int_{\Sigma}\mathrm{det}\, e\, \left(P_{m}\,\bar{\nabla} X^m+w_\alpha\,\bar{\nabla}\lambda^{\alpha}+\tilde{w}_{\til\alpha}\,\bar{\nabla}\til\lambda^{\til\alpha}+L^{\alpha}B_{\alpha}+\tilde{L}^{\til\alpha}\tilde{B}_{\til\alpha}+\bLambda_{\alpha}\lambda^{\alpha}+\tbLambda_{\til\alpha}\til\lambda^{\til\alpha}\right)\,.
\ee
Here $\bar{\nabla}=e^{J}_{+}\partial_{J}$ is the covariant derivative with $e^{J}_{\pm}$ the zweibein on the worldsheet $\Sigma$. Both $\lambda^{\alpha}$ and $\til\lambda^{\til\alpha}$ are pure spinors in ten-dimensions, meaning that they obey the algebraic condition
\begin{equation*}
 \lambda\gamma^{m}\lambda=0=\til\lambda \gamma^{m}\til\lambda\,,
\end{equation*}
and have only eleven independent complex components each. Recall that the space of pure spinors in ten-dimensions is given by $(\SO(10)/\U(5))\times\C^{*}$.

The Lagrange multipliers $L^{\alpha}, \tilde{L}^{\til\alpha}, \bLambda_{\alpha}$, and $\tbLambda_{\til\alpha}$ enforce the constraints
\be\label{const}
B_{\alpha}=-\frac{1}{2}P_{m}\,\gamma^{m}_{\alpha\beta}\lambda^{\beta}\,, \qquad \tilde{B}_{\til\alpha}= -\frac{1}{2}P_{m}\,\gamma^{m}_{\til\alpha \til\beta}\til\lambda^{\til\beta}\,,
\ee
and $\lambda^{\alpha}=0=\til\lambda^{\til\alpha}$, respectively. The latter will serve to eliminate potential divergences from the non-compact zero-mode integrations over the pure spinors, while the former are twistor-like constraints. Note that these twistor constraints are related to those appearing in the superstring by the replacement $P_{m}\leftrightarrow\partial X_{m}$~\cite{Berkovits:2014aia}. The pure spinor conditions on $\lambda,\til\lambda$ mean that $\bLambda$ and $\tbLambda$ are also pure.

The constraints $\{B, \tilde{B}, \lambda,\til\lambda\}$ and the various pure spinor conditions imply associated gauge invariances for the action \eqref{gf1} which must be fixed. Note that unlike the superstring studied in~\cite{Berkovits:2014aia}, the twistor constraints commute (i.e., $[B_{\alpha},B_{\beta}]=0$) so their associated gauge freedoms are simpler. In particular, the constraint $B_{\alpha}$ generates a symmetry of the action \eqref{gf1} given by
\be\label{gf2}
\delta X^{m}=\frac{\varepsilon}{2}(\lambda\gamma^{m} f)\,, \qquad \delta w_{\alpha}=-\frac{\varepsilon}{2}P_{m}(\gamma^{m}f)_{\alpha}\,, \qquad \delta L^{\alpha}=\varepsilon\,\bar{\nabla}f^{\alpha}\,,
\ee
where $f^{\alpha}\in\Pi\Omega^{0}(\Sigma)$ is the infinitesimal gauge transformation and $\varepsilon\in\Pi\Omega^{0}(\Sigma)$ is the associated parameter. The other constraint $\tilde{B}_{\til\alpha}$ induces the obvious analogue of \eqref{gf2} on the tilded variables.

The gauge invariances associated to the pure spinor condition are the same as in the superstring~\cite{Berkovits:2014aia}. For instance, the $\lambda^{\alpha}$ constraint implies the gauge invariance
\be\label{gf3}
\delta w_{\alpha}=\varepsilon\,g_{\alpha}\,, \qquad \delta\bLambda_{\alpha}=\varepsilon \left(\bar{\nabla}g_{\alpha}+\frac{(g\gamma^{m}\bar{\nabla}\bLambda)\,(\gamma_{m}\lambda)_\alpha}{2\,(\lambda\cdot\bLambda)}\right)\,,
\ee
for $g_{\alpha}\in\Pi\Omega^{0}(\Sigma, K_{\Sigma})$. Finally, we have the gauge freedom associated with the pure spinor constraint on $\lambda$ and $\til\lambda$:
\be\label{gf4}
\delta L^{\alpha}=c_{mn}(\gamma^{mn}\lambda)^{\alpha}\,,\qquad \delta\tilde{L}^{\til\alpha}=\tilde{c}_{mn}(\gamma^{mn}\til\lambda)^{\til\alpha}\,,
\ee
with $c_{mn},\tilde{c}_{mn}$ arrays of arbitrary parameters.

\medskip

At this point, it is worth noting that the constraints in \eqref{gf1} actually imply the two bosonic constraints which are gauged in the RNS-like formalism of this worldsheet theory~\cite{Mason:2013sva,Adamo:2013tsa}.  These are the Virasoro and Hamiltonian constraints, implemented by gauging the worldsheet currents
\begin{equation*}
 T=-P_{m}\nabla X^{m}\,,\qquad \cH=\eta^{mn} P_{m}\,P_{n}\,,
\end{equation*}
both of which take values in $\Omega^{0}(\Sigma,K_{\Sigma}^2)$. The latter of these is implied by the twistor constraints of the pure spinor action, since
\begin{equation*}
 \cH=-B_{\alpha}\frac{P_{n}(\gamma^{n}\bLambda)^{\alpha}}{\lambda\cdot\bLambda}\,.
\end{equation*}
The Virasoro constraint is more subtle as it is implied only by the twistor constraint in conjunction with $\lambda^{\alpha}$. In particular, the stress tensor of \eqref{gf1} is given by
\begin{equation*}
 T=B_{\alpha}\frac{\nabla X^{n}(\gamma_{n}\bLambda)^{\alpha}}{\lambda\cdot\bLambda}-\lambda^{\alpha}\frac{P_{m}\nabla X^{n}(\gamma^{\:m}_{n}\bLambda)_{\alpha}}{2\,\lambda\cdot\bLambda}+\nabla\lambda^{\alpha}\frac{(\lambda\gamma_{mn}w)(\gamma^{mn}\bLambda)_{\alpha}+2(\lambda\cdot w)\bLambda_{\alpha}}{8\, \lambda\cdot\bLambda}\,.
\end{equation*}
Hence, the twistor constraint $B_{\alpha}$ seems to be stronger in this `infinite tension' limit than its superstring analogue, which only implies the Virasoro constraint. This is likely related to the fact that the effective $b$-ghost \eqref{effbg} in our loop amplitude prescription obeys $\{Q,b\}=\cH$ rather than $\{Q,b\}=T$, while nevertheless providing what appears to be the correct measure for integrating over the moduli space. Of course, a more precise understanding of this relationship would be desirable.

\medskip

Since these constraints imply the Virasoro constraint, we are free to move to conformal gauge on the worldsheet, whereupon the action becomes:
\be\label{ungfact}
S=\frac{1}{2\pi}\int_{\Sigma} P_{m}\,\dbar X^m+w_\alpha\,\dbar\lambda^{\alpha}+\tilde{w}_{\til\alpha}\,\dbar\til\lambda^{\til\alpha}+L^{\alpha}B_{\alpha}+\tilde{L}^{\til\alpha}\tilde{B}_{\til\alpha}+\bLambda_{\alpha}\lambda^{\alpha}+\tbLambda_{\til\alpha}\til\lambda^{\til\alpha}\,.
\ee
We now fix the gauge redundancies \eqref{gf2}-\eqref{gf4} with the usual BRST procedure. In order to do this, we must work patchwise on the space of pure spinors. Consider the pure spinor $\lambda^{\alpha}$; the space of pure spinors has a natural open cover $\{U_{\alpha}\}$, where $U_{\alpha}$ is open subset of $(\SO(10)/\U(5))\times\C^{*}$ with the $\alpha^{\mathrm{th}}$-component of $\lambda$ is non-zero.

These patches can be specified by a choice of constant pure spinor of the opposite chirality, $\bar{\lambda}_{\alpha}$, and demanding that $\bar{\lambda}\cdot\lambda\neq0$.  With the manifestly $\U(5)$-covariant description of spinors in ten (Euclidean) dimensions, under which a spinor decomposes into the $(\mathbf{1}, \mathbf{10}, \bar{\mathbf{5}})$ representations of $\U(5)$, the pure spinor constraint is solved as (\textit{c.f.}, \cite{Berkovits:2000fe,Berkovits:2005bt})
\begin{equation*}
 \lambda^{\alpha}=(\lambda^{+},\,\lambda_{ij},\,\lambda^{i})=\left(\lambda^{+},\, \lambda_{ij},\,\frac{1}{8}\epsilon^{ijklm}\lambda_{jk}\lambda_{lm}\right)\,,
\end{equation*}
where $i,j=1,\ldots,5$ are $\U(5)$ vector indices. For instance, in the patch $U_{+}$ on which the component $\lambda^{+}\neq0$, one can specifying a pure spinor $\bar{\lambda}_{\alpha}=(\bar{\lambda}_{+}, 0, 0)$ for which $\bar{\lambda}\cdot\lambda\neq0$.

Now, on a given patch $U_{\beta}$, we can use \eqref{gf4} to fix eleven components of $L^{\alpha}$ by satisfying the condition $L\gamma^{mn}\bar{\lambda}=0$.  This leaves five free components in $L^{\alpha}$, which can be gauged to zero using \eqref{gf2}, with the transformation function $f^{\alpha}$ constrained by
\be\label{gff}
f\gamma^{mn}\bar{\lambda}=0\,.
\ee
The gauge freedom \eqref{gf3} can also be used to set $\bLambda_{\alpha}=\epsilon\bar{\lambda}_{\alpha}$, for some constant $\epsilon\neq0$. This imposes the constraint
\be\label{gfg}
g\gamma^{m}\bar{\lambda}=0\,,
\ee
on the transformation function $g_{\alpha}$.

Following the usual BRST-procedure to implement these gauge-fixings results in
\begin{multline*}
 S-\frac{1}{2\pi}\int_{\Sigma}M_{\alpha} L^{\alpha}-N^{\alpha}(\bLambda_{\alpha}-\epsilon\bar{\lambda}_{\alpha})+m_{\alpha}\dbar f^{\alpha}+n^{\alpha}\dbar g_{\alpha} \\
 -\tilde{M}_{\til\alpha}\tilde{L}^{\til\alpha}-\tilde{N}^{\til\alpha}(\tbLambda_{\til\alpha}-\til\epsilon\til{\bar{\lambda}}_{\til\alpha})+\tilde{m}_{\til\alpha}\dbar\tilde{f}^{\til\alpha}+\tilde{n}^{\til\alpha}\dbar\tilde{g}_{\til\alpha}\,,
\end{multline*}
where we have now included the tilded fields, with $m_{\alpha},\tilde{m}_{\til\alpha}\in\Pi\Omega^{0}(\Sigma, K_{\Sigma})$ and $n^{\alpha},\tilde{n}^{\til\alpha}\in\Pi\Omega^{0}(\Sigma)$ the antighosts for $f^{\alpha},\tilde{f}^{\til\alpha}$ and $g_{\alpha},\tilde{g}_{\til\alpha}$ respectively. The $M_{\alpha},\tilde{M}_{\til\alpha}\in\Omega^{0}(\Sigma,K_{\Sigma})$ and $N^{\alpha},\tilde{N}^{\til\alpha}\in\Omega^{0}(\Sigma)$ are the associated Nakanishi-Lautrup Lagrange multiplier fields. Furthermore, the BRST-operator can easily be calculated from the Noether procedure on \eqref{ungfact} using \eqref{gf2}, \eqref{gf3}:
\be\label{BRST1}
Q=\oint \lambda^{\alpha}\,g_{\alpha}+B_{\alpha}\,f^{\alpha}+\til\lambda^{\til\alpha}\,\tilde{g}_{\til\alpha}+\tilde{B}_{\til\alpha}\,\tilde{f}^{\til\alpha}\,.
\ee
The various constraints on the gauge-fixed fields also imply constraints on the anti-ghosts and Lagrange multipliers:
\be\label{agg}
 \lambda\gamma^{mn}m=0=\lambda\gamma^{mn}M\,, \qquad \lambda\gamma^{m}n=0=\lambda\gamma^{m}N\,,
\ee
along with their tilded partners.

Upon integrating out all Lagrange multipliers in the path integral, we are left with a gauge-fixed action
\begin{multline}\label{gfact1}
S=\frac{1}{2\pi}\int_{\Sigma}P_{m}\,\dbar X^m+w_\alpha\,\dbar\lambda^{\alpha}+\tilde{w}_{\til\alpha}\,\dbar\til\lambda^{\til\alpha}+m_{\alpha}\dbar f^{\alpha}+\tilde{m}_{\til\alpha}\dbar\tilde{f}^{\til\alpha}+n^{\alpha}\dbar g_{\alpha}+\tilde{n}^{\til\alpha}\dbar\tilde{g}_{\til\alpha}\\
+\epsilon\, \bar{\lambda}\cdot\lambda+\til\epsilon\,\til{\bar{\lambda}}\cdot\til\lambda\,.
\end{multline}
We now follow~\cite{Berkovits:2014aia} and define new fermionic fields:
\be\label{ferms}
\theta^{\alpha}\equiv f^{\alpha}+n^{\alpha}\,, \qquad p_{\alpha}\equiv g_{\alpha}+m_{\alpha}\,,
\ee
with $\tilde{\theta}^{\til\alpha}$ and $\tilde{p}_{\til\alpha}$ defined similarly. While each of $\{f,n,g,m\}$ are constrained by \eqref{gff}, \eqref{gfg}, \eqref{agg}, it is easy to see that $\theta^{\alpha}, p_{\alpha}$ are not. 

In particular, consider $\theta^{\alpha}$. The constraint \eqref{gff} implies that only five of the sixteen components of $f^{\alpha}$ and are undetermined; the remainder are fixed in terms of $\bar{\lambda}_{\alpha}$. Eleven of the sixteen components of $n^{\alpha}$ are independent due to \eqref{agg}, and these are precisely complimentary to the free components of $f^{\alpha}$. In terms of the $\U(5)$-decomposition, the independent components of $f^{\alpha}$ lie in the $\bar{\mathbf{5}}$ representation, while those of $n^{\alpha}$ are in the $\mathbf{1}$ and $\mathbf{10}$ representations. A similar statement obtains for the constituents of $p_{\alpha}$. 

So the new fermionic variables are totally unconstrained, and hence independent of the choice of patch $U_{\beta}$ in pure spinor space on which we have performed the gauge fixing. This means that the terms proportional to $\bar{\lambda}, \til{\bar{\lambda}}$ can be eliminated from the action \eqref{gfact1} by sending $\epsilon,\til\epsilon\rightarrow 0$, leaving us with
\be\label{gfact2}
S=\frac{1}{2\pi}\int_{\Sigma}P_{m}\,\dbar X^m+w_\alpha\,\dbar\lambda^{\alpha}+\tilde{w}_{\til\alpha}\,\dbar\til\lambda^{\til\alpha}+p_{\alpha}\,\dbar\theta^{\alpha}+\tilde{p}_{\til\alpha}\,\dbar\til\theta^{\til\alpha}\,,
\ee
and the BRST-charge
\be\label{BRST2}
Q=\oint \lambda^{\alpha}\left(p_\alpha-\frac{1}{2}P_{m}\gamma^{m}_{\alpha\beta}\theta^{\beta}\right)+\til\lambda^{\til\alpha}\left(\tilde{p}_{\til\alpha}-\frac{1}{2}P_{m}\gamma^{m}_{\til\alpha \til\beta}\til\theta^{\beta}\right) =\oint \lambda^{\alpha}\,d_{\alpha}+\til\lambda^{\til\alpha}\,\tilde{d}_{\til\alpha} \,.
\ee
These are precisely the action and BRST-charges of the minimal model~\cite{Berkovits:2013xba}, given by \eqref{min1}, \eqref{minBRST}. The non-minimal variables can also be obtained from a gauge-fixing, in direct analogy with the process for the superstring~\cite{Berkovits:2014aia}.

\bibliography{PS2}
\bibliographystyle{JHEP}

\end{document}